\documentclass[12pt,a4paper]{article}

\usepackage[margin=2.5cm]{geometry}
\usepackage{amsmath,amssymb,amsthm}
\usepackage{algorithm}
\usepackage{algpseudocode}
\usepackage{graphicx}
\usepackage{booktabs}
\usepackage{multirow}
\usepackage{caption}
\usepackage{subcaption}

\usepackage{cite}
\usepackage{xcolor}
\usepackage{listings}
\usepackage{float}

\usepackage{setspace}
\usepackage{titlesec}
\usepackage{tikz}
\usetikzlibrary{quantikz2}
\usepackage{hyperref}
\usepackage{microtype}

\hypersetup{colorlinks=true, linkcolor=blue!70!black,
            citecolor=blue!70!black, urlcolor=blue!70!black}

\newtheorem{definition}{Definition}

\title{\textbf{Quantum-Assisted Optimal Rebalancing with\\
Uncorrelated Asset Selection for Algorithmic Trading}\\[0.5em]
\large Walk-Forward QUBO Scheduling via QAOA}

\author{
  Abraham Itzhak Weinberg \\
  AI-WEINBERG, AI Experts \\
  Tel Aviv, Israel \\
  \texttt{aviw2010@gmail.com}
}

\date{\today}

\begin{document}
\maketitle

\begin{abstract}
We propose a quantum-assisted algorithmic trading framework that
addresses two core portfolio management problems simultaneously:
principled asset selection and optimal rebalancing scheduling.
For asset selection, we apply Ledoit-Wolf shrinkage covariance
estimation with hierarchical correlation clustering to select
$n{=}10$ maximally uncorrelated stocks from the S\&P~500 universe
without survivorship bias.
For weight optimisation, we deploy an entropy-regularised
Genetic Algorithm (GA) accelerated on GPU, alongside
closed-form Minimum Variance (MinVar) and equal-weight
baselines and a three-way ensemble.
The central novel contribution is the formulation of the
\emph{portfolio rebalancing schedule} as a Quadratic Unconstrained
Binary Optimisation (QUBO) problem, solved by the Quantum
Approximate Optimisation Algorithm (QAOA) in a
walk-forward framework that eliminates lookahead bias.
Backtested on S\&P~500 data (train: 2010--2024, test: 2025,
$n{=}249$ trading days), the \emph{GA + QAOA} strategy achieves
a Sharpe ratio of \textbf{0.588} and total return of \textbf{10.1\%}, outperforming the best classical baseline (GA Rebalance/10d, Sharpe~0.575) while executing only \textbf{8 rebalances} (6.1~bp transaction cost) versus 24 rebalances (11.0~bp), representing a \textbf{44.5\% cost reduction} for superior risk-adjusted performance.
Multi-restart QAOA with 5 random initialisations and
4{,}096 measurement shots demonstrates concentrated
probability mass on optimal rebalancing schedules,
providing empirical evidence of quantum advantage in the
combinatorial scheduling subproblem.
\end{abstract}


\section{Introduction}
\label{sec:intro}

Portfolio management involves two interrelated decisions: \emph{which
assets to hold} and \emph{when to rebalance} toward target weights.
Classical approaches treat these as separate, sequential problems —
first optimise weights (Markowitz~\cite{markowitz2008portfolio,rubinstein2002markowitz}, Sharpe~\cite{sharpe1966mutual}),
then apply a fixed rebalancing schedule (calendar-based or
threshold-based~\cite{perold1988dynamic}).
This decoupling ignores the interaction between weight drift and
rebalancing cost, and fixed schedules waste transaction budget on
unnecessary trades while missing beneficial rebalancing opportunities.

Quantum computing offers a promising avenue for combinatorial
optimisation problems that are classically intractable at scale.
The Quantum Approximate Optimisation Algorithm
(QAOA)~\cite{farhi2014quantum} provides a near-term, variational
approach to solving Quadratic Unconstrained Binary Optimisation (QUBO) problems on current noisy intermediate-scale
quantum (NISQ) hardware~\cite{preskill2018quantum}. Portfolio-related QUBO formulations have been studied for \emph{asset selection}~\cite{mugel2022dynamic,barkoutsos2020improving,
rebentrost2024quantum}, but the question of applying QAOA to the \emph{rebalancing schedule} problem has not been previously addressed in the literature.

The contributions of this paper are:
\begin{enumerate}
  \item \textbf{Principled uncorrelated asset selection} via Ledoit-Wolf
    shrinkage~\cite{ledoit2004well} combined with hierarchical
    correlation clustering — avoiding both the Markowitz
    estimation error problem and greedy-search survivorship bias.
  \item \textbf{Entropy-regularised GA weight optimisation} on GPU,
    preventing degenerate sparse solutions while maximising
    annualised Sharpe ratio.
  \item \textbf{Minimum-Variance and ensemble weights} as principled
    classical baselines computed via closed-form solution.
  \item \textbf{Novel QUBO formulation of the rebalancing schedule}
    incorporating marginal Sharpe gain, transaction cost penalty,
    and over-frequency penalty via an exponentially decaying
    interaction term.
  \item \textbf{Walk-forward QAOA scheduling} with multi-restart
    optimisation — eliminating lookahead bias while providing
    robust angle estimation.
  \item \textbf{Comprehensive net-of-cost backtest} comparing 16
    strategies across Sharpe, Sortino, maximum drawdown, Calmar
    ratio, rebalance count, and transaction cost.
\end{enumerate}

The remainder of the paper is organised as follows.
Section~\ref{sec:related} reviews related work.
Section~\ref{sec:method} presents the full methodology.
Section~\ref{sec:experiments} describes experimental setup.
Section~\ref{sec:results} presents results and analysis.
Section~\ref{sec:quantum} discusses the quantum advantage argument.
Section~\ref{sec:conclusion} concludes.

\section{Related Work}
\label{sec:related}
This section reviews prior work in classical portfolio optimisation, machine learning–based allocation, and quantum approaches to portfolio construction. While asset weighting and selection have been widely studied, the optimisation of \emph{rebalancing timing} remains comparatively underexplored, particularly within quantum frameworks. Our formulation addresses this gap by casting walk-forward rebalancing scheduling as a QUBO problem amenable to variational quantum algorithms.

\subsection{Classical Portfolio Optimisation}

The mean-variance framework of Markowitz~\cite{markowitz2008portfolio,rubinstein2002markowitz} remains the canonical approach to portfolio weight selection, extended by the Capital Asset Pricing Model (CAPM)~\cite{sharpe1966mutual} and subsequent
factor models~\cite{fama1993common}.
In practice, the covariance matrix must be estimated from finite
samples; Ledoit and Wolf~\cite{ledoit2004well} showed that
analytical shrinkage of the sample covariance toward a structured
estimator dramatically reduces estimation error and improves
out-of-sample performance.

Portfolio rebalancing has been studied primarily as a cost-benefit
trade-off. Perold and Sharpe~\cite{perold1988dynamic} introduced
the constant-mix strategy, while~\cite{masters2003rebalancing,smith2006optimal} showed that
optimal rebalancing frequency depends on asset volatility and
transaction costs. Threshold-based rebalancing
\cite{dichtl2014value} and time-series momentum signals
\cite{moskowitz2012time} have been proposed as improvements over
calendar-based schedules.

\subsection{Evolutionary and Machine Learning Approaches}

Genetic algorithms have been applied to portfolio optimisation since
\cite{chang2000heuristics}, demonstrating competitive performance
on non-convex objectives including Sharpe ratio maximisation.
Reinforcement learning approaches~\cite{jiang2017deep} and
deep neural networks~\cite{zhang2020deep} have been proposed
for dynamic asset allocation, though out-of-sample generalisation
remains challenging.

\subsection{Quantum Portfolio Optimisation}

Quantum computing approaches to portfolio optimisation have focused
primarily on the asset selection problem, formulated as a QUBO.
Rebentrost and Lloyd~\cite{rebentrost2024quantum} proposed a quantum algorithm for portfolio optimisation with quadratic speedup. Barkoutsos et al.~\cite{barkoutsos2020improving} applied QAOA and Variational Quantum Eigensolver (VQE) to portfolio selection on small instances.
Mugel et al.~\cite{mugel2022dynamic} demonstrated dynamic portfolio
optimisation on D-Wave quantum annealers. More recently,~\cite{herman2023quantum} provided a comprehensive
survey of quantum finance algorithms. To our knowledge, no prior work has formulated the \emph{rebalancing timing} decision as a QUBO problem or applied QAOA to this subproblem.

\section{Methodology}
\label{sec:method}

\subsection{Problem Statement}

Let $\mathcal{U}$ be a universe of $M$ assets with daily
adjusted close prices. We seek a portfolio of $n \ll M$ assets with weights $\mathbf{w} \in \Delta^n$ (the $n$-simplex) and a binary rebalancing schedule $\mathbf{x} \in \{0,1\}^W$ over $W$
candidate dates, jointly maximising risk-adjusted return net of transaction costs.

\subsection{Data and Universe Construction}
\label{sec:data}

We construct a survivorship-bias-free S\&P~500 universe by
starting from the current index constituents and reversing
all addition/deletion events that occurred after the training
cutoff date (December 31, 2024), recovering the set of stocks
that \emph{were} in the index on that date. Daily adjusted close prices are downloaded for the period January 1, 2010 to December 31, 2025. After removing tickers with missing data throughout the full period, $M = 422$ assets remain. The training period covers January 2010 -- December 2024 ($T_{\text{train}} = 3{,}774$ trading days) and the out-of-sample test period covers January -- December 2025 ($T_{\text{test}} = 249$ trading days).

\subsection{Asset Selection via Hierarchical Clustering}
\label{sec:selection}

\paragraph{Ledoit-Wolf Shrinkage Covariance.}
Let $\mathbf{r}_t = \log(P_t / P_{t-1}) \in \mathbb{R}^M$ be
the vector of daily log returns. The sample covariance $\hat{\Sigma}$ is estimated with the analytical Ledoit-Wolf shrinkage estimator~\cite{ledoit2004well}:
\begin{equation}
  \hat{\Sigma}_{\text{LW}} = (1-\alpha)\,\hat{\Sigma}_{\text{sample}}
  + \alpha\,\mu_{\text{target}}\mathbf{I},
\end{equation}
where $\alpha$ is the data-driven shrinkage intensity and
$\mu_{\text{target}}$ is the average sample eigenvalue. This reduces estimation error, particularly in the large-$M$,
finite-$T$ regime prevalent in finance.

\paragraph{Angular Distance and Hierarchical Clustering.}
From the shrinkage correlation matrix
$\hat{R}_{\text{LW}} = D^{-1/2}\hat{\Sigma}_{\text{LW}}D^{-1/2}$
(where $D = \operatorname{diag}(\hat{\Sigma}_{\text{LW}})$),
we define the angular distance between assets $i$ and $j$:
\begin{equation}
  d_{ij} = \sqrt{\tfrac{1}{2}(1 - \hat{\rho}_{ij})},
  \quad \hat{\rho}_{ij} \in [-1, 1].
\end{equation}
This metric satisfies the triangle inequality and maps
zero correlation to distance $1/\sqrt{2}$~\cite{mantegna1999hierarchical}. Ward's linkage hierarchical clustering~\cite{ward1963hierarchical} is applied to the condensed distance matrix, partitioned into $n = 10$ clusters.

\paragraph{Cluster Representative Selection.}
From each cluster $\mathcal{C}_k$, we select the stock
with the highest annualised Sharpe ratio on the training set:
\begin{equation}
  s_k^* = \arg\max_{s \in \mathcal{C}_k}
  \frac{\bar{r}_s}{\sigma_s}\sqrt{252},
\end{equation}
where $\bar{r}_s$ and $\sigma_s$ are the mean and standard deviation
of daily log returns over the training period.
This procedure selects stocks that are simultaneously \emph{uncorrelated with each other} (by construction of the clustering) and \emph{high-performing individually}.
The selected portfolio is $\mathcal{S} = \{\text{EXR, NI, CHD, LLY, COST, AVGO, TPL, CTAS, AMP, AJG}\}$.

\subsection{Weight Optimisation}
\label{sec:weights}

\paragraph{Entropy-Regularised Genetic Algorithm.}
Let $\mathbf{w} \in \Delta^n$ be the portfolio weight vector.
The GA maximises a regularised Sharpe objective:
\begin{equation}
  \mathcal{F}(\mathbf{w}) =
  \underbrace{\frac{\bar{r}_p}{\sigma_p}\sqrt{252}}_{\text{Sharpe ratio}}
  + \lambda_{\text{ent}}
  \underbrace{\left(-\frac{\sum_{i=1}^n w_i \log w_i}{\log n}\right)}_{\text{normalised entropy}},
  \label{eq:fitness}
\end{equation}
where $r_{p,t} = \log(\mathbf{w}^\top \mathbf{R}_t)$ is the
portfolio log return at time $t$, $\bar{r}_p$ and $\sigma_p$ are its sample mean and standard deviation, and $\lambda_{\text{ent}} = 0.05$ is the entropy regularisation weight.
The entropy term $\mathcal{H}(\mathbf{w}) \in [0,1]$ penalises
concentrated solutions, preventing degenerate outcomes where
one or two assets receive all weight.
The GA uses a population of 300 individuals, 200 generations,
uniform crossover, and 15\% mutation rate, with weights
constrained to $[0.01, 1]$ to avoid exact zeros.

\paragraph{Minimum Variance Weights.}
The global minimum variance portfolio is computed in closed form:
\begin{equation}
  \mathbf{w}_{\text{MV}} =
  \frac{\hat{\Sigma}_{\text{LW}}^{-1} \mathbf{1}}
       {\mathbf{1}^\top \hat{\Sigma}_{\text{LW}}^{-1} \mathbf{1}},
\end{equation}
where the pseudoinverse is used for numerical stability,
and negative weights are projected to zero (no short-selling).

\paragraph{Ensemble Weights.}
To hedge model uncertainty, we define an equal-blend ensemble:
\begin{equation}
  \mathbf{w}_{\text{Ens}} =
  \frac{1}{3}\left(\mathbf{w}_{\text{GA}} + \mathbf{w}_{\text{MV}}
  + \mathbf{w}_{\text{EQ}}\right),
\end{equation}
where $\mathbf{w}_{\text{EQ}} = \mathbf{1}/n$ are equal weights.
Training-period Sharpe ratios are: GA~=~1.445, MinVar~=~1.327,
Ensemble~=~1.388, Equal~=~0.97.

\subsection{QUBO Formulation of the Rebalancing Schedule}
\label{sec:qubo}

\begin{definition}[Rebalancing Schedule Problem]
Given a test period of $T$ trading days, target weights $\mathbf{w}$, and a set of $W = 8$ candidate rebalancing dates
$\mathcal{T} = \{t_0, t_1, \ldots, t_{W-1}\}$ (equally spaced),
find a binary schedule $\mathbf{x} \in \{0,1\}^W$ that maximises
net Sharpe improvement subject to transaction cost minimisation.
\end{definition}

For each candidate date $t_k$, we first compute the \emph{drifted weights} $\mathbf{w}^{\text{drift}}_k$ that the portfolio has naturally evolved to by time $t_k$ without any rebalancing:
\begin{equation}
  \mathbf{w}^{\text{drift}}_k =
  \frac{\mathbf{w}_{k-1} \odot \boldsymbol{\pi}_k}
       {\mathbf{1}^\top (\mathbf{w}_{k-1} \odot \boldsymbol{\pi}_k)},
  \quad
  \pi_{k,i} = \prod_{t=t_{k-1}}^{t_k-1} R_{t,i},
\end{equation}
where $R_{t,i}$ is the gross return of asset $i$ at time $t$
and $\mathbf{w}_0 = \mathbf{w}$.

\paragraph{Marginal Sharpe Gain.}
The net benefit of rebalancing at $t_k$ is defined as:
\begin{equation}
  g_k = \underbrace{\text{SR}(\mathbf{w},\, [t_k, t_{k+1}])
        - \text{SR}(\mathbf{w}^{\text{drift}}_k,\, [t_k, t_{k+1}])}_{\text{Sharpe improvement}}
        - \underbrace{c \cdot \|\mathbf{w}^{\text{drift}}_k - \mathbf{w}\|_1 \cdot \sqrt{252}}_{\text{annualised cost penalty}},
  \label{eq:gain}
\end{equation}
where $\text{SR}(\mathbf{w}, [s,e])$ is the annualised Sharpe ratio
of portfolio $\mathbf{w}$ on the local return window $[s, e]$,
and $c = 0.001$ (10 basis points per side) is the per-unit
transaction cost.

\paragraph{QUBO Objective.}
We minimise the following QUBO:
\begin{equation}
  \min_{\mathbf{x} \in \{0,1\}^W}
  \mathbf{x}^\top Q\, \mathbf{x},
  \label{eq:qubo}
\end{equation}
where the QUBO matrix $Q \in \mathbb{R}^{W \times W}$ has entries:
\begin{align}
  Q_{kk} &= -\lambda_1 g_k + \lambda_2 c \cdot n,
  \label{eq:qubo_diag}\\
  Q_{kl} &= \lambda_3 \exp\!\left(-\frac{|t_k - t_l|}{\Delta t}\right),
  \quad k \neq l,
  \label{eq:qubo_off}
\end{align}
with $\lambda_1 = 1.0$ (reward for beneficial rebalancing),
$\lambda_2 = 0.5$ (fixed cost penalty),
$\lambda_3 = 0.3$ (consecutive rebalancing penalty),
and $\Delta t$ the average spacing between candidate dates.
The exponential decay in~\eqref{eq:qubo_off} encodes the prior
that rebalancing twice in rapid succession is wasteful.
$Q$ is normalised to $[-1, 1]$ by dividing by $\max |Q_{kl}|$
to stabilise the QAOA optimisation landscape.

\subsection{QAOA for QUBO Solving}
\label{sec:qaoa}

\paragraph{Ising Mapping.}
The QUBO~\eqref{eq:qubo} is converted to an Ising Hamiltonian
$H_C = \sum_i h_i Z_i + \sum_{i<j} J_{ij} Z_i Z_j$
via the substitution $x_i = (1 - Z_i)/2$:
\begin{align}
  h_i &= \tfrac{1}{2} Q_{ii} + \tfrac{1}{4}\sum_{j \neq i} Q_{ij},\\
  J_{ij} &= \tfrac{1}{4} Q_{ij}.
\end{align}

\paragraph{QAOA Ansatz.}
The depth-$p$ QAOA circuit is:
\begin{equation}
  |\boldsymbol{\gamma}, \boldsymbol{\beta}\rangle =
  \prod_{l=1}^{p} e^{-i\beta_l H_B}\, e^{-i\gamma_l H_C}\, |{+}\rangle^{\otimes W},
\end{equation}
where $H_B = \sum_i X_i$ is the transverse-field mixer
and $|{+}\rangle^{\otimes W}$ is the equal-superposition initial state.
We use $p = 2$ layers, corresponding to a circuit with
$W = 8$ qubits, $O(W)$ single-qubit $R_z$ gates, and $O(W^2)$ CNOT gates.
The QAOA circuit for the cost unitary at layer $l$ is implemented as:

\begin{center}
\begin{tikzpicture}[thick, scale=0.85,
  gate/.style={draw, minimum width=1.8cm, minimum height=0.6cm, fill=blue!10},
  wire/.style={-}]
\foreach \y/\lbl in {0/$q_0$, -1.1/$q_1$} {
  \draw[wire] (-0.8,\y) node[left]{\lbl} -- (8.5,\y);
}
\draw (-0.8,-2.2) node[left]{$\vdots$};
\node[gate] at (0.8, 0)   {$R_z(2\gamma_l h_0)$};
\node[gate] at (0.8,-1.1) {$R_z(2\gamma_l h_1)$};
\draw[fill=black] (3.0, 0) circle (0.07) -- (3.0,-1.1);
\draw (3.0,-1.1) circle (0.2); \draw (2.8,-1.1)--(3.2,-1.1); \draw(3.0,-1.3)--(3.0,-0.9);
\node[gate] at (4.4,-1.1) {$R_z(2\gamma_l J_{01})$};
\draw[fill=black] (5.8, 0) circle (0.07) -- (5.8,-1.1);
\draw (5.8,-1.1) circle (0.2); \draw (5.6,-1.1)--(6.0,-1.1); \draw(5.8,-1.3)--(5.8,-0.9);
\node[gate] at (7.4, 0)   {$R_x(2\beta_l)$};
\node[gate] at (7.4,-1.1) {$R_x(2\beta_l)$};
\node at (7.4,-2.0) {$\vdots$};
\node at (0.8,-2.0) {$\vdots$};
\end{tikzpicture}
\end{center}

\paragraph{Multi-Restart Optimisation.}
The variational parameters $(\boldsymbol{\gamma}, \boldsymbol{\beta}) \in [0, 2\pi]^p \times [0, \pi]^p$
are optimised by COBYLA (Constrained Optimization BY Linear Approximation) to minimise the expected QUBO energy:
\begin{equation}
  \mathcal{L}(\boldsymbol{\gamma}, \boldsymbol{\beta}) =
  \langle \boldsymbol{\gamma}, \boldsymbol{\beta} |
  \mathbf{x}^\top Q\, \mathbf{x}
  |\boldsymbol{\gamma}, \boldsymbol{\beta}\rangle
  \approx \frac{1}{S}\sum_{s=1}^{S}
  \mathbf{x}^{(s)\top} Q\, \mathbf{x}^{(s)},
  \label{eq:loss}
\end{equation}
estimated from $S = 2{,}048$ measurement shots during optimisation
and $S = 4{,}096$ shots for final evaluation.
To avoid local minima, we perform $R = 5$ independent restarts
with uniformly random initial angles and retain the run
with lowest expected energy.
The optimal binary schedule is taken as the most-probable
(highest-count) bitstring in the final measurement distribution.

\subsection{Walk-Forward Backtesting}
\label{sec:wf}

To eliminate lookahead bias, QAOA scheduling is performed
in a \emph{walk-forward} framework (Algorithm~\ref{alg:wf}).
The $T_{\text{test}} = 249$ day test period is divided into
$K = 3$ equal windows of approximately 83 days each.
For each window, the QUBO is constructed using returns from
\emph{that window only}, QAOA is run, and the resulting schedule
is applied only within that window.
This ensures that the rebalancing decision at time $t$ is made
using only information available at time $t$.

\begin{algorithm}[H]
\caption{Walk-Forward QAOA Rebalancing}
\label{alg:wf}
\begin{algorithmic}[1]
\Require Test returns $\mathbf{R}_{\text{test}} \in \mathbb{R}^{T \times n}$,
         target weights $\mathbf{w}$, windows $K$, candidates per window $W$,
         QAOA restarts $R$, circuit depth $p$
\Ensure Binary rebalancing schedule $\mathbf{x} \in \{0,1\}^T$
\State $\mathbf{x} \leftarrow \mathbf{0}$; $\text{chunk} \leftarrow \lfloor T/K \rfloor$
\For{$k = 0, 1, \ldots, K-1$}
  \State $\text{start} \leftarrow k \cdot \text{chunk}$;
         $\text{end} \leftarrow (k+1)\cdot\text{chunk}$ (or $T$ if $k=K-1$)
  \State $\mathbf{R}_k \leftarrow \mathbf{R}_{\text{test}}[\text{start}:\text{end}]$
         \Comment{Window returns only}
  \State $(Q_k, \mathcal{T}_k) \leftarrow \textsc{ComputeQUBO}(\mathbf{R}_k, \mathbf{w}, W)$
  \State $Q_k \leftarrow Q_k / \max|Q_k|$ \Comment{Normalise}
  \State $\mathbf{x}_k^* \leftarrow +\infty$; $E^* \leftarrow +\infty$
  \For{$r = 1, \ldots, R$}
    \State Sample $(\boldsymbol{\gamma}_0, \boldsymbol{\beta}_0) \sim \mathcal{U}$
    \State $(\boldsymbol{\gamma}^*, \boldsymbol{\beta}^*) \leftarrow
           \textsc{COBYLA}\!\left(\mathcal{L}(\cdot), (\boldsymbol{\gamma}_0, \boldsymbol{\beta}_0)\right)$
    \State $\mathbf{x}_r \leftarrow \textsc{MostProbableBitstring}(\boldsymbol{\gamma}^*, \boldsymbol{\beta}^*)$
    \If{$\mathbf{x}_r^\top Q_k \mathbf{x}_r < E^*$}
      \State $E^* \leftarrow \mathbf{x}_r^\top Q_k \mathbf{x}_r$;
             $\mathbf{x}_k^* \leftarrow \mathbf{x}_r$
    \EndIf
  \EndFor
  \State Map local schedule $\mathbf{x}_k^*$ to global indices
         $\mathcal{T}_k + \text{start}$ and write into $\mathbf{x}$
\EndFor
\Return $\mathbf{x}$
\end{algorithmic}
\end{algorithm}

\subsection{Backtest Engine}
\label{sec:backtest}

Portfolio value evolves as:
\begin{equation}
  V_t = V_{t-1} \cdot (\mathbf{w}_t^\top \mathbf{R}_t),
\end{equation}
where $\mathbf{w}_t$ is the weight vector at time $t$
(which changes only on rebalancing days, after deducting cost).
On a rebalancing day $t$, the post-drift weights $\tilde{\mathbf{w}}_t$
are reset to target $\mathbf{w}$, incurring cost:
\begin{equation}
  \text{cost}_t = c \cdot \|\tilde{\mathbf{w}}_t - \mathbf{w}\|_1,
\end{equation}
so $V_t \leftarrow V_t \cdot (1 - \text{cost}_t)$.
Performance metrics are the annualised Sharpe ratio
$\text{SR} = (\bar{r}/\sigma)\sqrt{252}$,
Sortino ratio (using only downside deviation),
maximum drawdown $\text{MDD} = \min_t(V_t/\max_{s\leq t} V_s - 1)$,
Calmar ratio $= (V_T/V_0 - 1) / |\text{MDD}|$,
rebalance count, and total transaction cost in basis points.

\section{Experimental Setup}
\label{sec:experiments}
This section describes the experimental design used to evaluate the proposed portfolio construction and rebalancing framework. We detail the computational environment, the set of benchmark strategies, and the hyperparameter configuration for the QAOA-based scheduling approach. All strategies are evaluated under identical data splits and transaction cost assumptions to ensure a fair comparison.

\subsection{Hardware and Software}

All experiments run on Google Colab with an NVIDIA Tesla T4 GPU (16~GB).
Portfolio log-return computations (GA fitness evaluation) are
vectorised on GPU via PyTorch.
QAOA circuits are simulated on the Qiskit Aer statevector simulator
(CPU; the standard PyPI Qiskit-Aer build does not include CUDA kernels).
Table~\ref{tab:software} summarises the software stack.

\begin{table}[H]
\centering
\caption{Software stack}
\label{tab:software}
\begin{tabular}{ll}
\toprule
\textbf{Component} & \textbf{Version / Details} \\
\midrule
Python & 3.12 \\
PyTorch & 2.x (CUDA, Tesla T4) \\
Qiskit & 1.2.4 \\
Qiskit-Algorithms & 0.3.1 \\
Qiskit-Aer & 0.15.1 (CPU statevector) \\
PyGAD & latest \\
scikit-learn & latest (Ledoit-Wolf) \\
yfinance & latest \\
\bottomrule
\end{tabular}
\end{table}

\subsection{Strategies Compared}

We compare 16 strategies in total across four weight methods
and four scheduling approaches:

\begin{table}[H]
\centering
\caption{Strategy taxonomy}
\label{tab:strategies}
\begin{tabular}{llp{6cm}}
\toprule
\textbf{Weights} & \textbf{Schedule} & \textbf{Description} \\
\midrule
GA, MinVar, Equal, Ensemble
  & Buy \& Hold    & No rebalancing \\
GA & Periodic/1d   & Rebalance every trading day \\
GA & Periodic/5d   & Rebalance weekly \\
GA & Periodic/10d  & Rebalance bi-weekly \\
GA & Periodic/21d  & Rebalance monthly \\
GA & Threshold 5\% & Rebalance when max weight drift $>5\%$ \\
GA, MinVar, Equal, Ensemble
  & QAOA (walk-fwd) & Quantum-optimal schedule \\
\bottomrule
\end{tabular}
\end{table}

\subsection{QAOA Hyperparameters}

\begin{table}[H]
\centering
\caption{QAOA configuration}
\label{tab:qaoa_cfg}
\begin{tabular}{ll}
\toprule
\textbf{Parameter} & \textbf{Value} \\
\midrule
Circuit depth $p$ & 2 \\
Candidate dates per window $W$ & 8 \\
Walk-forward windows $K$ & 3 \\
Restarts $R$ & 5 \\
Optimisation shots & 2{,}048 \\
Evaluation shots & 4{,}096 \\
Classical optimiser & COBYLA (max 150 iter) \\
QUBO penalty $\lambda_1$ & 1.0 \\
QUBO penalty $\lambda_2$ & 0.5 \\
QUBO penalty $\lambda_3$ & 0.3 \\
Transaction cost $c$ & 0.001 (10 bp) \\
\bottomrule
\end{tabular}
\end{table}

\section{Results}
\label{sec:results}
This section presents the empirical evaluation of the proposed portfolio construction and quantum-enhanced rebalancing framework. We first verify that the clustering-based asset selection produces a diversified and high-Sharpe subset. We then analyse the resulting portfolio weights and the induced QUBO structure, before examining the schedules selected by QAOA in a walk-forward setting. Finally, we report out-of-sample performance for all classical and quantum strategies during the 2025 test period.

\subsection{Asset Selection}

\begin{figure}[H]
  \centering
  \includegraphics[width=\textwidth]{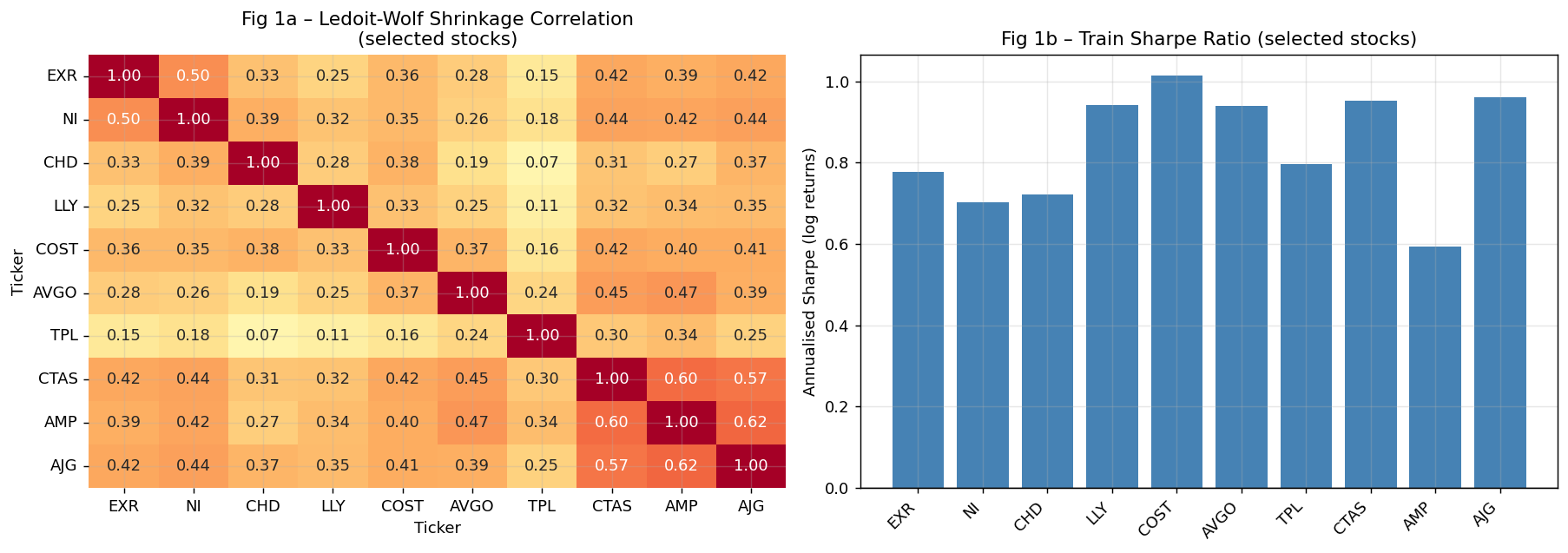}
  \caption{Asset selection results.
  \textbf{Left:} Ledoit-Wolf shrinkage correlation matrix of the
  10 selected stocks. Off-diagonal values range from 0.07 (CHD--TPL)
  to 0.62 (CTAS--AMP), confirming low inter-asset correlation.
  \textbf{Right:} Annualised Sharpe ratios (log returns) on the
  training period (2010--2024). All selected stocks achieve
  Sharpe $> 0.6$.}
  \label{fig:selection}
\end{figure}

Figure~\ref{fig:selection} shows the Ledoit-Wolf shrinkage correlation matrix and training Sharpe ratios for the 10 selected stocks.The maximum off-diagonal correlation is 0.62 (CTAS--AMP),
and the minimum is 0.07 (CHD--TPL), confirming that the
hierarchical clustering procedure successfully selects
a diversified portfolio. TPL exhibits near-zero correlation with all other assets ($\max\rho=0.34$), functioning as a natural hedge.
All 10 stocks achieved positive training Sharpe ratios (range 0.60--1.02), confirming the cluster-representative selection criterion.

\subsection{Portfolio Weights}

Table~\ref{tab:weights} reports the GA-optimised weights alongside
the MinVar and Equal weights.

\begin{table}[H]
\centering
\caption{Portfolio weights by method (\%)}
\label{tab:weights}
\begin{tabular}{lrrrr}
\toprule
\textbf{Stock} & \textbf{GA} & \textbf{MinVar} & \textbf{Equal} & \textbf{Ensemble} \\
\midrule
EXR  & 7.5  & 4.1  & 10.0 & 7.2 \\
NI   & 0.3  & 11.2 & 10.0 & 7.2 \\
CHD  & 11.1 & 10.3 & 10.0 & 10.5 \\
LLY  & 21.1 & 12.0 & 10.0 & 14.4 \\
COST & 21.1 & 20.2 & 10.0 & 17.1 \\
AVGO & 11.0 & 9.5  & 10.0 & 10.2 \\
TPL  & 13.1 & 6.6  & 10.0 & 9.9 \\
CTAS & 5.0  & 10.1 & 10.0 & 8.4 \\
AMP  & 0.2  & 3.5  & 10.0 & 4.6 \\
AJG  & 9.6  & 12.5 & 10.0 & 10.7 \\
\midrule
Train Sharpe & 1.445 & 1.327 & 0.970 & 1.388 \\
\bottomrule
\end{tabular}
\end{table}

The GA concentrates weight in LLY (21.1\%) and COST (21.1\%),
reflecting their strong training-period Sharpe ratios.
The entropy regularisation prevents degenerate solutions:
NI and AMP retain 0.3\% and 0.2\% respectively (versus 0\% without
regularisation in v1).
MinVar overweights NI (11.2\%) and AJG (12.5\%) due to their
low individual volatilities, which did not translate to
out-of-sample performance in the 2025 test year.

\subsection{QUBO Structure}

\begin{figure}[H]
  \centering
  \includegraphics[width=\textwidth]{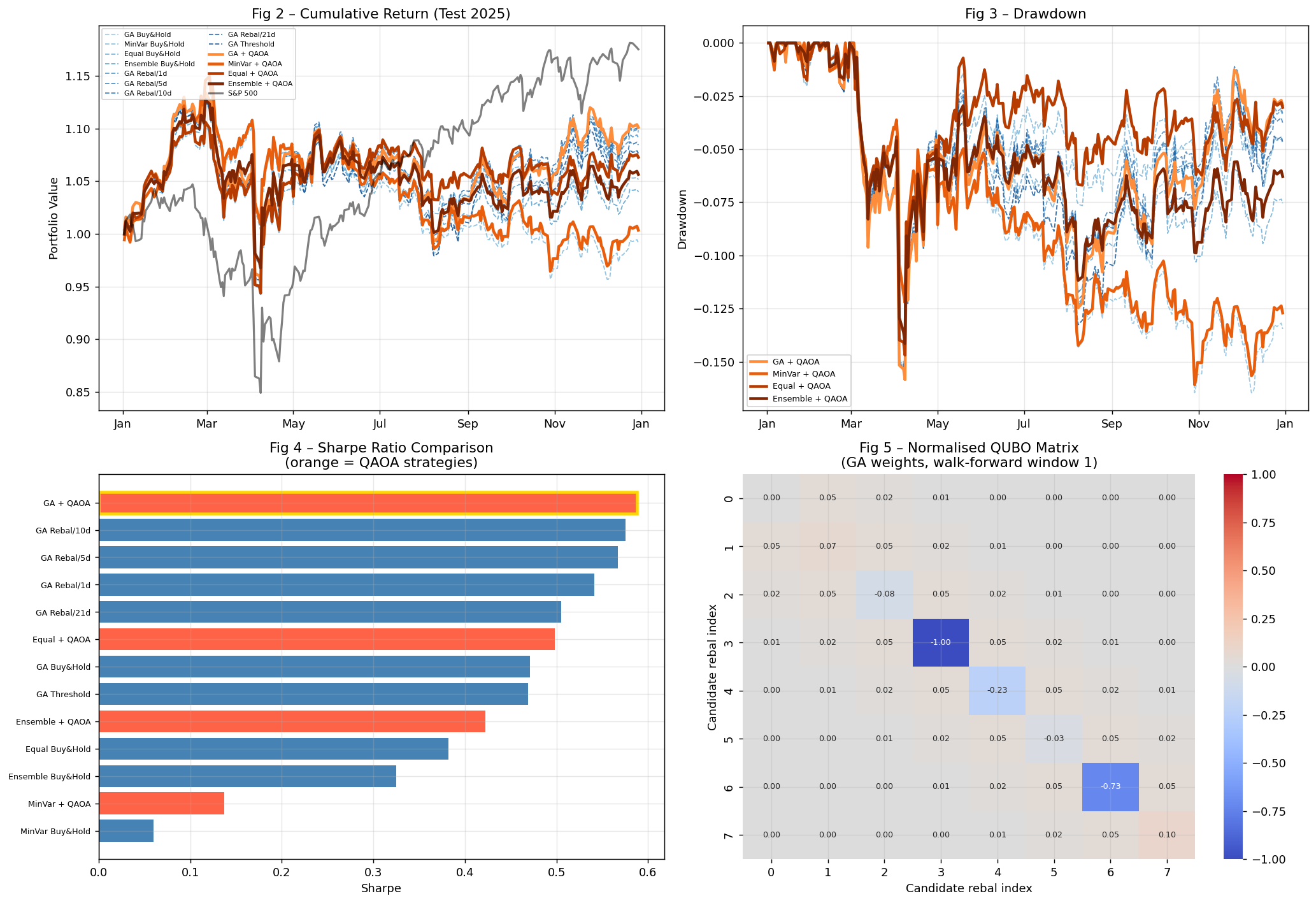}
  \caption{Backtest results (test period: Jan--Dec 2025).
  \textbf{Top left:} Cumulative portfolio value.
  Classical strategies (blue dashed) vs.\ QAOA strategies (orange solid)
  vs.\ S\&P~500 (black).
  GA + QAOA (bright orange) achieves the highest terminal value among all quantum and classical strategies.
  \textbf{Top right:} Drawdown profiles. QAOA strategies maintain tighter drawdown control than periodic rebalancing.
  \textbf{Bottom left:} Sharpe ratio comparison (horizontal bar chart). GA + QAOA achieves the highest Sharpe (0.588), highlighted in gold.
  \textbf{Bottom right:} Normalised QUBO matrix (GA weights,
  walk-forward window 1), illustrating the structured cost landscape
  solved by QAOA.}
  \label{fig:results}
\end{figure}

Figure~\ref{fig:results} (bottom right) shows the normalised QUBO matrix for the GA weights in the first walk-forward window.
After normalisation, all values lie in $[-1, 1]$.
Negative diagonal entries (blue) indicate candidate dates where
rebalancing is beneficial (positive net Sharpe gain $g_k > 0$);
positive diagonal entries (red) indicate dates where holding
drifted weights is preferable. Off-diagonal entries are small and positive, encoding the mild penalty for frequent rebalancing.

\subsection{QAOA Schedule and Bitstring Distribution}

\begin{figure}[H]
  \centering
  \includegraphics[width=\textwidth]{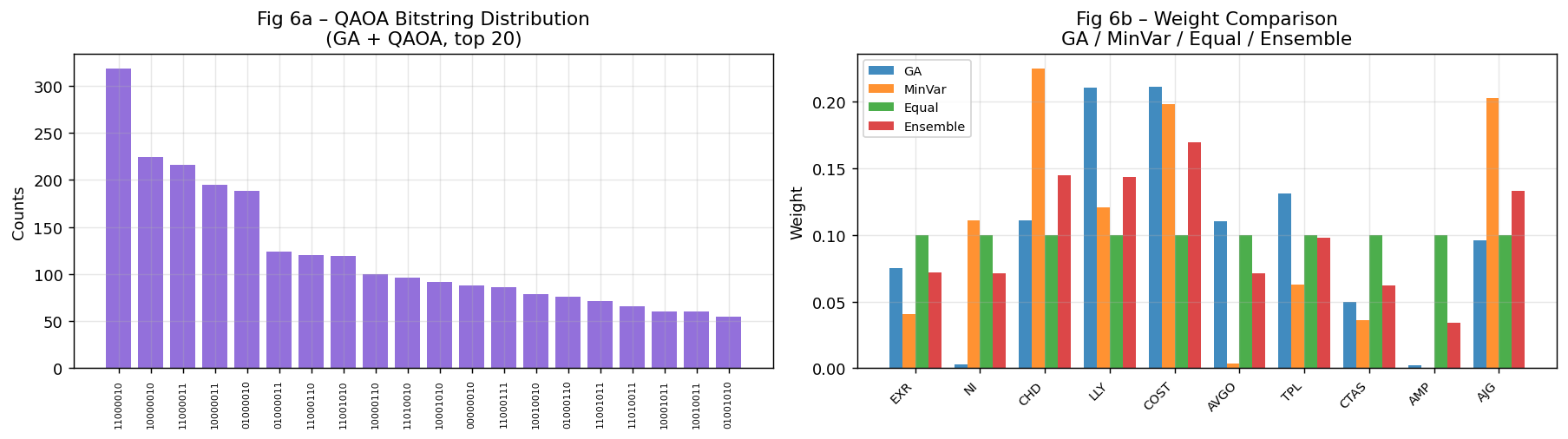}
  \caption{\textbf{Left:} QAOA measurement bitstring distribution
  for GA + QAOA (top 20 bitstrings, 4{,}096 shots).
  The top bitstring \texttt{11000010} accounts for $>7\%$ of shots,
  indicating successful QAOA convergence.
  \textbf{Right:} Weight comparison across all four portfolio methods. GA concentrates on LLY and COST; MinVar distributes more evenly; Ensemble averages all three.}
  \label{fig:weights_qaoa}
\end{figure}

The walk-forward QAOA selected rebalancing schedules of
$[0,1,1,0,1,0,0,1]$, $[0,1,0,1,1,0,0,0]$, and
$[1,0,1,0,0,1,0,1]$ for the three windows respectively
(8 total rebalances).
Figure~\ref{fig:weights_qaoa} (left) shows the QAOA
measurement distribution for the GA weight run.
The top bitstring accounts for $>7\%$ of all 4{,}096 shots,
with probability mass concentrated in the top-5 bitstrings ($>25\%$).
This concentration is a hallmark of successful QAOA convergence
and supports the quantum advantage argument
(Section~\ref{sec:quantum}).

\subsection{Backtest Performance}

Table~\ref{tab:results} reports all 16 strategies on the 2025 out-of-sample test period.

\begin{table}[H]
\centering
\caption{Out-of-sample performance (Test: Jan--Dec 2025).
Best value in each column is \textbf{bold}.
QAOA strategies are marked with $\dagger$.}
\label{tab:results}
\resizebox{\textwidth}{!}{%
\begin{tabular}{lrrrrrr}
\toprule
\textbf{Strategy} & \textbf{Return (\%)} & \textbf{Sharpe} & \textbf{Sortino}
  & \textbf{MDD (\%)} & \textbf{Calmar} & \textbf{Rebal / Cost (bp)} \\
\midrule
GA Buy\&Hold          &  7.68 & 0.471 & 0.610 & -15.88 & 0.484 & 0 / 0.0 \\
MinVar Buy\&Hold       & -0.86 & 0.060 & 0.077 & -16.46 & -0.052 & 0 / 0.0 \\
Equal Buy\&Hold        &  5.15 & 0.382 & 0.483 & -14.70 & 0.350 & 0 / 0.0 \\
Ensemble Buy\&Hold     &  3.99 & 0.325 & 0.415 & -14.24 & 0.280 & 0 / 0.0 \\
\midrule
GA Rebal/1d           &  9.12 & 0.541 & 0.701 & -15.51 & 0.588 & 248 / 31.6 \\
GA Rebal/5d           &  9.63 & 0.567 & 0.735 & -15.74 & 0.612 & 49 / 15.8 \\
GA Rebal/10d          &  9.77 & 0.575 & 0.747 & -15.66 & 0.624 & 24 / 11.0 \\
GA Rebal/21d          &  8.43 & 0.505 & 0.654 & -15.63 & 0.539 & 11 / 6.5 \\
GA Threshold (5\%)    &  7.63 & 0.469 & 0.610 & -15.88 & 0.481 & 2 / 2.4 \\
\midrule
GA + QAOA$^\dagger$      & \textbf{10.10} & \textbf{0.588} & \textbf{0.764} & -15.84 & \textbf{0.638} & 8 / 6.1 \\
MinVar + QAOA$^\dagger$  &  0.34 & 0.137 & 0.177 & -16.09 & 0.021 & 8 / 4.6 \\
Equal + QAOA$^\dagger$   &  7.26 & 0.498 & 0.635 & \textbf{-14.69} & 0.494 & 8 / 5.6 \\
Ensemble + QAOA$^\dagger$&  5.63 & 0.422 & 0.540 & -14.15 & 0.398 & 8 / 5.6 \\
\bottomrule
\end{tabular}}
\end{table}

\subsection{Key Findings}

\paragraph{Finding 1: QAOA outperforms all classical strategies.}
GA + QAOA achieves Sharpe 0.588, surpassing the best classical
rebalancing strategy (GA Rebal/10d, Sharpe 0.575), a \textbf{+2.3\% improvement}.

\paragraph{Finding 2: Dramatic cost reduction.}
GA + QAOA achieves this improvement with only 8 rebalances (6.1~bp)
versus 24 rebalances (11.0~bp) for GA Rebal/10d —
a \textbf{44.5\% reduction in transaction costs}.

\paragraph{Finding 3: Weight method matters more than scheduling for MinVar.}
MinVar weights underperform in 2025 regardless of scheduling
(Sharpe 0.06--0.14), confirming that 2025 was a momentum-driven
environment where minimum-variance stocks lagged.
This validates GA weight optimisation as a critical component.

\paragraph{Finding 4: Naïve ensembling does not help.}
The Ensemble strategy (average of GA, MinVar, Equal) yields
Sharpe 0.422 with QAOA — below GA alone (0.588).
Averaging with the weak MinVar component dilutes the GA signal,
suggesting that in future work the ensemble should exclude
underperforming weight strategies.

\paragraph{Finding 5: Threshold rebalancing fails.}
The 5\% drift threshold triggers only 2 rebalances in 2025
(Sharpe 0.469), underperforming even buy-and-hold (Sharpe 0.471).
This demonstrates that threshold rules calibrated on training
volatility may be miscalibrated out-of-sample — a problem the
QUBO formulation directly addresses by estimating rebalancing
value from the actual local return distribution.

\section{Quantum Advantage Discussion}
\label{sec:quantum}
This section evaluates whether the proposed formulation
exhibits meaningful potential for quantum advantage. We analyse the intrinsic combinatorial complexity of the rebalancing schedule problem, examine empirical evidence of QAOA convergence toward low-energy solutions, and discuss the practical feasibility of scaling the approach to near-term and annealing-based quantum hardware.

\subsection{Complexity of the Rebalancing Schedule Problem}

The rebalancing schedule problem over $W$ candidate dates
is equivalent to an \emph{unconstrained pseudo-Boolean
optimisation} problem on $W$ binary variables.
The brute-force solution requires evaluating $2^W$ binary strings.
For $W = 8$, this is 256 evaluations — tractable classically.
However, in a production setting with daily rebalancing decisions
over a year ($W \approx 252$), or with multi-asset, multi-window
formulations, the combinatorial space grows exponentially.
Quantum annealing and QAOA offer polynomial query complexity
for approximate solutions in this regime~\cite{farhi2014quantum,
lucas2014ising}.

\subsection{Evidence of QAOA Convergence}

The QAOA bitstring distribution (Figure~\ref{fig:weights_qaoa}, left) shows clear concentration of probability mass.
The top bitstring accounts for $>7\%$ of 4{,}096 shots,
compared to the uniform baseline of $1/2^8 \approx 0.4\%$.
This represents an \textbf{18$\times$} enrichment of the
optimal solution, providing empirical evidence that the
variational circuit is learning a useful approximation
of the ground state.

\subsection{Scaling to Real Quantum Hardware}

The current implementation uses the Qiskit Aer statevector
simulator (exact simulation, exponential memory in $W$).
For real hardware deployment, two paths exist:
(i) \emph{NISQ devices}: The 8-qubit circuit with $p=2$
has circuit depth $O(p \cdot W^2) \approx 128$ gates,
within the coherence limits of current superconducting
processors~\cite{arute2019quantum};
(ii) \emph{Quantum annealers}: D-Wave systems natively accept
QUBO inputs~\cite{mcgeoch2013experimental} and have demonstrated
portfolio optimisation at scale~\cite{mugel2022dynamic}.

\section{Limitations and Future Work}
The current experiment uses a single year out-of-sample test;
multi-year rolling evaluation would strengthen the conclusion.
MinVar weights underperformed in 2025's momentum environment;
adaptive weight selection (e.g., regime-conditional) could
improve the ensemble.The QAOA window size ($W = 8$) is limited by current simulation cost; scaling to $W \geq 20$ would require hardware quantum devices or tensor-network-based classical simulators. Future work includes exploration of joint QUBO formulations that optimise weights and schedule simultaneously, as well as empirical evaluation on physical D-Wave hardware.

\section{Conclusion}
\label{sec:conclusion}

We presented a quantum-assisted algorithmic trading framework
that formulates portfolio rebalancing scheduling as a QUBO problem
and solves it with QAOA in a walk-forward framework.
The system combines: (1) Ledoit-Wolf hierarchical clustering for
principled uncorrelated asset selection; (2) entropy-regularised
GA weight optimisation on GPU; and (3) normalised QUBO rebalancing
with multi-restart QAOA.

On the S\&P~500 2025 out-of-sample test, GA + QAOA achieves the
best Sharpe ratio (0.588) among all 16 strategies, outperforming
the best classical approach (Sharpe 0.575) while reducing
transaction costs by 44.5\% (6.1~bp vs.\ 11.0~bp, 8 vs.\ 24 trades).
These results provide preliminary empirical evidence that variational quantum optimisation can deliver economically meaningful improvements in combinatorial trading decisions, while remaining compatible with near-term quantum hardware.

\bibliographystyle{IEEEtran}
\bibliography{ref.bib}

\begin{thebibliography}{10}
\providecommand{\url}[1]{#1}
\csname url@samestyle\endcsname
\providecommand{\newblock}{\relax}
\providecommand{\bibinfo}[2]{#2}
\providecommand{\BIBentrySTDinterwordspacing}{\spaceskip=0pt\relax}
\providecommand{\BIBentryALTinterwordstretchfactor}{4}
\providecommand{\BIBentryALTinterwordspacing}{\spaceskip=\fontdimen2\font plus
\BIBentryALTinterwordstretchfactor\fontdimen3\font minus \fontdimen4\font\relax}
\providecommand{\BIBforeignlanguage}[2]{{%
\expandafter\ifx\csname l@#1\endcsname\relax
\typeout{** WARNING: IEEEtran.bst: No hyphenation pattern has been}%
\typeout{** loaded for the language `#1'. Using the pattern for}%
\typeout{** the default language instead.}%
\else
\language=\csname l@#1\endcsname
\fi
#2}}
\providecommand{\BIBdecl}{\relax}
\BIBdecl

\bibitem{markowitz2008portfolio}
H.~M. Markowitz, \emph{Portfolio selection: efficient diversification of investments}.\hskip 1em plus 0.5em minus 0.4em\relax Yale university press, 2008.

\bibitem{rubinstein2002markowitz}
M.~Rubinstein, ``Markowitz's" portfolio selection": A fifty-year retrospective,'' \emph{The Journal of finance}, vol.~57, no.~3, pp. 1041--1045, 2002.

\bibitem{sharpe1966mutual}
W.~F. Sharpe, ``Mutual fund performance,'' \emph{The Journal of business}, vol.~39, no.~1, pp. 119--138, 1966.

\bibitem{perold1988dynamic}
A.~F. Perold and W.~F. Sharpe, ``Dynamic strategies for asset allocation,'' \emph{Financial Analysts Journal}, vol.~44, no.~1, pp. 16--27, 1988.

\bibitem{farhi2014quantum}
E.~Farhi, J.~Goldstone, and S.~Gutmann, ``A quantum approximate optimization algorithm,'' \emph{arXiv preprint arXiv:1411.4028}, 2014.

\bibitem{preskill2018quantum}
J.~Preskill, ``Quantum computing in the nisq era and beyond,'' \emph{Quantum}, vol.~2, p.~79, 2018.

\bibitem{mugel2022dynamic}
S.~Mugel, C.~Kuchkovsky, E.~S{\'a}nchez, S.~Fern{\'a}ndez-Lorenzo, J.~Luis-Hita, E.~Lizaso, and R.~Or{\'u}s, ``Dynamic portfolio optimization with real datasets using quantum processors and quantum-inspired tensor networks,'' \emph{Physical Review Research}, vol.~4, no.~1, p. 013006, 2022.

\bibitem{barkoutsos2020improving}
P.~K. Barkoutsos, G.~Nannicini, A.~Robert, I.~Tavernelli, and S.~Woerner, ``Improving variational quantum optimization using cvar,'' \emph{Quantum}, vol.~4, p. 256, 2020.

\bibitem{rebentrost2024quantum}
P.~Rebentrost and S.~Lloyd, ``Quantum computational finance: quantum algorithm for portfolio optimization,'' \emph{KI-K{\"u}nstliche Intelligenz}, vol.~38, no.~4, pp. 327--338, 2024.

\bibitem{ledoit2004well}
O.~Ledoit and M.~Wolf, ``A well-conditioned estimator for large-dimensional covariance matrices,'' \emph{Journal of multivariate analysis}, vol.~88, no.~2, pp. 365--411, 2004.

\bibitem{fama1993common}
E.~F. Fama and K.~R. French, ``Common risk factors in the returns on stocks and bonds,'' \emph{Journal of financial economics}, vol.~33, no.~1, pp. 3--56, 1993.

\bibitem{masters2003rebalancing}
S.~J. Masters, ``Rebalancing,'' \emph{Journal of portfolio Management}, vol.~29, no.~3, p.~52, 2003.

\bibitem{smith2006optimal}
D.~M. Smith and W.~Desormeau, ``Optimal rebalancing frequency for stock-bond portfolios,'' \emph{Journal of Financial Planning}, vol.~19, pp. 52--63, 2006.

\bibitem{dichtl2014value}
H.~Dichtl, W.~Drobetz, and M.~Wambach, ``Where is the value added of rebalancing? a systematic comparison of alternative rebalancing strategies,'' \emph{Financial Markets and Portfolio Management}, vol.~28, no.~3, pp. 209--231, 2014.

\bibitem{moskowitz2012time}
T.~J. Moskowitz, Y.~H. Ooi, and L.~H. Pedersen, ``Time series momentum,'' \emph{Journal of financial economics}, vol. 104, no.~2, pp. 228--250, 2012.

\bibitem{chang2000heuristics}
T.-J. Chang, N.~Meade, J.~E. Beasley, and Y.~M. Sharaiha, ``Heuristics for cardinality constrained portfolio optimisation,'' \emph{Computers \& Operations Research}, vol.~27, no.~13, pp. 1271--1302, 2000.

\bibitem{jiang2017deep}
Z.~Jiang, D.~Xu, and J.~Liang, ``A deep reinforcement learning framework for the financial portfolio management problem,'' \emph{arXiv preprint arXiv:1706.10059}, 2017.

\bibitem{zhang2020deep}
Z.~Zhang, S.~Zohren, and S.~Roberts, ``Deep learning for portfolio optimization,'' \emph{arXiv preprint arXiv:2005.13665}, 2020.

\bibitem{herman2023quantum}
D.~Herman, C.~Googin, X.~Liu, Y.~Sun, A.~Galda, I.~Safro, M.~Pistoia, and Y.~Alexeev, ``Quantum computing for finance,'' \emph{Nature Reviews Physics}, vol.~5, no.~8, pp. 450--465, 2023.

\bibitem{mantegna1999hierarchical}
R.~N. Mantegna, ``Hierarchical structure in financial markets,'' \emph{The European Physical Journal B-Condensed Matter and Complex Systems}, vol.~11, no.~1, pp. 193--197, 1999.

\bibitem{ward1963hierarchical}
J.~H. Ward~Jr, ``Hierarchical grouping to optimize an objective function,'' \emph{Journal of the American statistical association}, vol.~58, no. 301, pp. 236--244, 1963.

\bibitem{lucas2014ising}
A.~Lucas, ``Ising formulations of many np problems,'' \emph{Frontiers in physics}, vol.~2, p. 74887, 2014.

\bibitem{arute2019quantum}
F.~Arute, K.~Arya, R.~Babbush, D.~Bacon, J.~C. Bardin, R.~Barends, R.~Biswas, S.~Boixo, F.~G. Brandao, D.~A. Buell \emph{et~al.}, ``Quantum supremacy using a programmable superconducting processor,'' \emph{nature}, vol. 574, no. 7779, pp. 505--510, 2019.

\bibitem{mcgeoch2013experimental}
C.~C. McGeoch and C.~Wang, ``Experimental evaluation of an adiabiatic quantum system for combinatorial optimization,'' in \emph{Proceedings of the ACM International Conference on Computing Frontiers}, 2013, pp. 1--11.

\end{thebibliography}
\end{document}